\documentclass[prb,twocolumn,showpacs,amsmath,amssymb]{revtex4-1}
\usepackage{graphicx,amsmath,bm, braket, txfonts, color}
\begin{document}

%\title{Electronic detection of a microwave quantum state}
%\title{Dynamical Coulomb Blockade theory of a quantum conductor coupled to a prepared microwave cavity}

\title{Photon cross-correlations emitted by a Josephson junction in two microwave cavities}
\author{Mircea Trif}
\affiliation{Laboratoire de Physique des Solides, Universit\'e Paris-Sud, 91405 Orsay, France}
%\author{Fabien Portier}
%\affiliation{CEA-Service Physique de l'Etat Condens\'ee, 91190 Gif sur Yvette, France}
\author{Pascal Simon}
\affiliation{Laboratoire de Physique des Solides, Universit\'e Paris-Sud, 91405 Orsay, France}
 
 \date{\today}

\begin{abstract}
{We study a voltage biased  Josephson junction coupled to two resonators of incommensurate frequencies. Using a density approach to analyze the cavity fields and an input-output description to analyze the emitted photonic fluxes and their correlation functions, we have shown, both for infinite and finite bandwidth detectors,  that the emitted radiation is non-classical in the sense that the correlators violates Cauchy-Schwarz inequalities. We have  also studied the time dependence of the photonic correlations and showed that their line-width becomes narrower with the increase of the emission rate approaching from below the threshold limit.
}
\end{abstract}

\maketitle

\section{Introduction}

In atomic cavity quantum electrodynamics (cQED), the electric dipole moment of  an isolated atom  interacts with the
vacuum state electric field of the cavity. 
The quantum nature of the electromagnetic field can give rise to coherent Rabi oscillations between the atom and the cavity
provided the relaxation and decoherence rate are smaller than the Rabi frequency \cite{Haroche2001}. Similar physics can be observed when the atom
is replaced by a two-level solid atate system, often called an artificial atom, capacitively coupled to a single mode of the microwave cavity \cite{Walraff2004,Girvin2008}.
Such paradigmatic systems can be described by a Hamiltonian $H=H_{\rm sys}+H_{\rm cavity}+H_{\rm int}$ where $H_{\rm sys}$ denotes the Hamiltonian of the atom or artificial atom. The interaction of light and matter is encoded in $H_{\rm int}$ while $H_{\rm cavity}$ describes the cavity Hamiltonian. 

When the system under study is characterized by a finite number of degrees of freedom, the physics is rather well understood. 
However, replacing the atom by a mesoscopic conductor with open Fermi reservoirs leads to a dramatically different physics as the whole Fermi sea is affected by the coupling to the cavity. 
Such
 setups composed of a quantum conductor coupled to a cavity have been realized experimentally, both in experiments using metallic tunnel junctions \cite{Holst1994,Deblock2010}, as well as in more recent experiments with high-$Q$ microwave cavities coupled to either quantum dots \cite{Frey2012,Basset2013,Petta2014} or carbon nanotubes \cite{Delbecq2011,Basset2012,Viennot2014}. 

%Furthermore, mesoscopic conductors can be driven out-of-equilibrium. What may be seen as a source of complexity turns out to be an advantage since a direct measurement of their transport properties allows an access to the effects of the environment.
For an electric conductor, the dimensionless parameter that encodes the light-matter interaction is given by the ratio between the  impedance of the environment and the quantum of resistance $R_{\rm K}=h/e^2$. It can therefore be controlled and significantly increased by modulating the impedance of the environment. Under such conditions, one can potentially analyze the two sides of our system, namely the electronic one by standard transport measurements or the optical one by measuring the power emitted in the cavity. 

The influence of  an electromagnetic environment on the electronic transport properties of a coherent conductor has been studied experimentally rather extensively in the past twenty five years. Such a phenomenon is  well-understood for quantum conductors in the tunneling regime and is called dynamical Coulomb Blockade (DCB).\cite{odintsov,nazarov89,devoret90,girvin90} This  occurs when a quantum  coherent conductor, such as a tunnel junction, is inserted into a circuit.
 When an electron tunnels through a junction, this entails sudden voltage variations at the edge of the environment which excite its
electromagnetic modes. The backaction of the circuit then affects the charge transfer through the junction and 
its conductance is reduced in a nonlinear way.\cite{IN}
In this respect, the microwave cavity acts as structured environment that can absorb  photons (or emit at finite temperature). DCB has been experimentally probed both for non-resonant environments \cite{Delsing1989,Geerligs1989,Cleland1990,Pierre2007} as well as for environments formed by resonators, \cite{Holst1994,Deblock2010,Hofheinz,Rimberg2011,Portier2014,Rimberg2014} with excellent 
theoretical agreement.

% Maybe a sentence beyond the tunneling regime (Pierre et al., Finkelstein et al. also some theoretical work including mine !)
%
%In a normal metal tunnel junction, biased at voltage $V$, the energy $eV$ of a tunnelling electron can be dissipated both into quasiparticle excitations in the electrodes and into photons. At low temperature energy conservation forbids tunnelling  processes emitting photons with total energy higher than $eV$. This suppression reduces the conductance at low bias voltage. \cite{IN}
%

\begin{figure}[t]
\begin{center}
\includegraphics[width=0.99\linewidth]{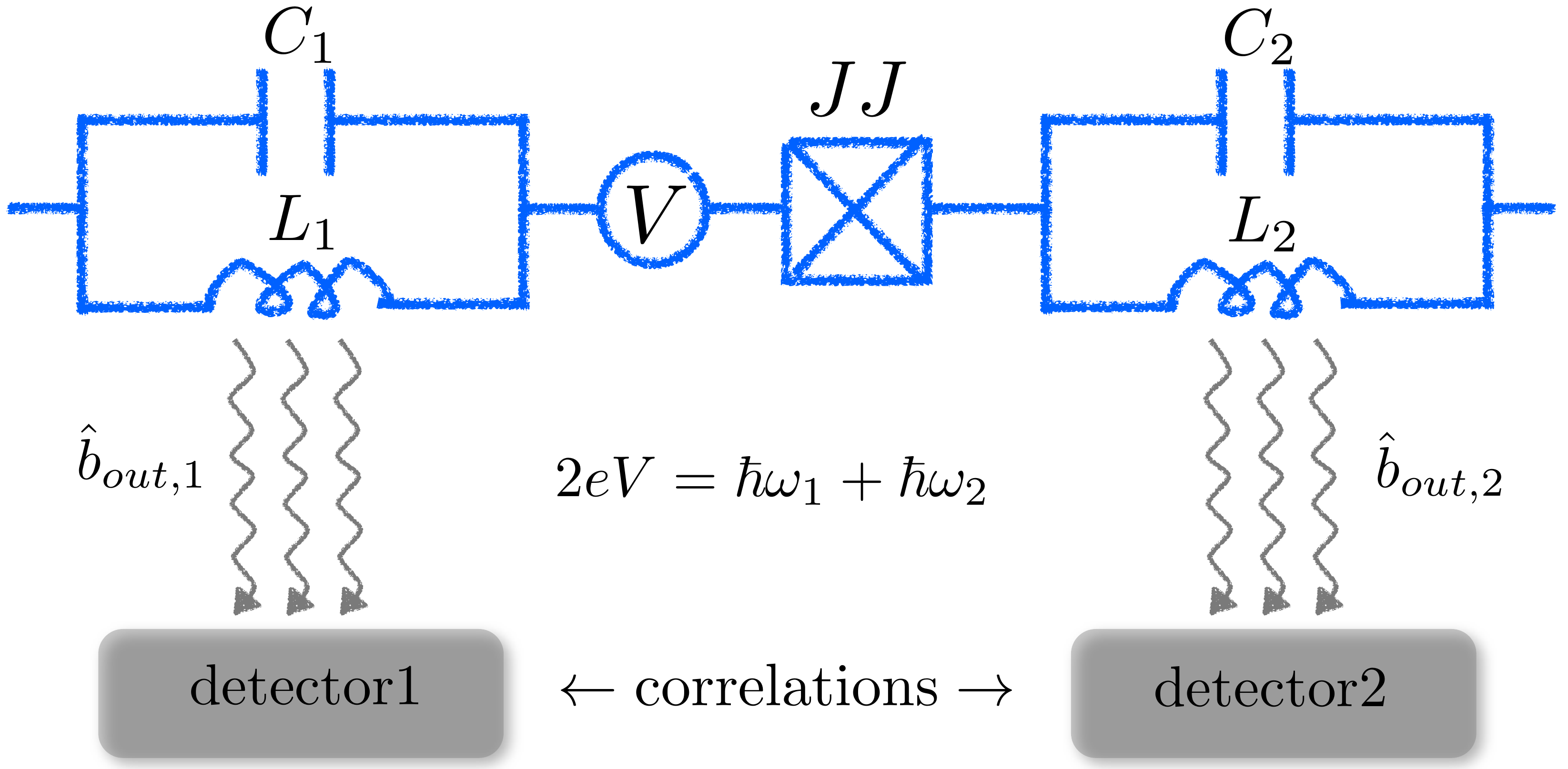}
\caption{The sketch of the  Josephson junction system coupled to two resonators. The $L_{1(2)}$ and $C_{1(2)}$ elements stand for the left (right) inductance and capacitance, respectively, and $V$ is the voltage biased applied over the JJ. The oscillators frequency is such that $2eV=\hbar\omega_1+\hbar\omega_2$. Also,  $\hat {b}_{out,1(2)}$ stand for the emitted radiation by the left (right) oscillator and collected by the left (right) detector. This radiation is expected to show strong quantum-mechanical correlations (see text).}
\label{sketch_JJ}
\end{center}
\end{figure}

For a Josephson junction (JJ), DCB effects are more dramatic since at bias  voltages  smaller  than $2\Delta/e$, with $\Delta$ the  superconducting gap,
quasi-particles cannot be excited.  Therefore the only possible channel to dissipate the  energy of tunnelling Cooper pairs is
to transform it into photons emitted in the electromagnetic environment.\cite{Averin90,IN} 
In a recent experiment,\cite{Hofheinz} Hofheinz {\it et al.}  have observed and characterized the radiation associated to the flow of Cooper pair connected to a microwave resonator. In particular, it was evidenced  two photon processes, for which a single Cooper pair tunnelling through the junction emits two photons into the environment.\cite{Hofheinz} 
Such two-photons processes which are triggered by  tunneling through a quantum conductor are very interesting because they can carry non-classical correlations and may therefore offer a source of pairs of correlated photons which could be very useful for quantum communications.\cite{Gisin2002}
The microwave radiation emitted by a Josephson junction coupled to a single resonator has been theoretically studied recently in various regimes. \cite{Padurariu2012,Lep2013,Ankerhold2013,Armour2013} 
In particular, it has  been shown by Leppagankas {\it et al.} \cite{Lep2013} that the radiation emitted by  the two photon-processes is non-classical, and that it violates a classical Cauchy-Schwarz inequality for two mode power cross-correlated fluctuations. 
Here we theoretically extend these considerations by analyzing the radiation emitted by  a Josephson junction coupled to two resonators of different cavities $\omega_1\neq\omega_2$. The device we study is sketched in Fig. 1. We analyze the correlations between the photons emitted by the JJ in the two resonators at bias matching the condition $2eV=\hbar\omega_1+\hbar\omega_2$, but also the power correlations of the microwave fields in the corresponding transmission lines using an input-output formalism.

The paper is organized as follows. In Sec.~\ref{sec2} we introduce
the setup and the system Hamiltonian of the Josephson junction
coupled to two LC oscillators. In Sec.~\ref{sec3} we introduce the
reduced density matrix describing the coupled system and calculate
various observables, such as the average photon number
and second order coherence functions. In Sec.~\ref{sec4} we utilize
the input-output theory to characterize the emitted light by the
two oscillators, for both infinite and finite bandwidths of the
detectors. We also calculate the time-dependent second order
coherence functions and calculate the resulting Fano factor of
the radiation. Finally, in Sec.~\ref{sec5} we end up with some conclusions
and outlook.

% The basic idea is that the probability of emitting a photon in each of the resonators during the observation time  is higher that the geometric mean of the probability of emitting two photons in either one of them. 

\section{Model Hamiltonian}
\label{sec2}

In the following we will describe the proposed  model and the associated Hamiltonian. In Fig.~\ref{sketch_JJ} we show a schematics of the voltage biased Josephson junction (JJ) coupled to two distinct $LC$ oscillators which, themselves, are coupled to a  transmission line that  can be used to probe each of oscillator independently. The model Hamiltonian describing the entire setup, including the external lines (which in Fig.~\ref{sketch_JJ} quantify the emitted radiation) reads:    
\begin{align}
H_{\rm tot}&=\sum_{i=1,2}(H_{\rm res}^i+H_{\rm env}^i+H_{\rm r-e}^i)+H_{J}\,,
\end{align}
being the sum of the resonators Hamiltonians, the environment (the external transmission lines), their coupling to the corresponding resonator, and the Josephson junction Hamiltonian, respectively. Specifically, 
\begin{equation}
H_{\rm res}^i=\frac{q_i^2}{2C_i}+\left(\frac{h}{2e}\right)^2\frac{1}{2L_i}\phi_i^2\,,
\end{equation}
is the Hamiltonian of the $i$th ($i=1,2$) resonator, with $\phi_i$ and $q_i$ being the phase and charge operators acting on the capacitance  $C_i$ and inductance $L_i$, respectively \cite{Lep2013,Ankerhold2013,Armour2013}. The second term  $H_{\rm env}^i=\sum_{q}\omega_{i,q}a_{q,i}^\dagger a_{q,i}$ is the Hamiltonian of the $i$th environment, with $a^{\dagger}_{q,i}$ ($a_{q,i}$) being the photon creation (annihilation) operator for the photons in that environment, with $\omega_{q,i}$ and $q$ being the energy and their wave-vector, respectively. The third term represents the  coupling between the oscillator and the environment reads:
\begin{equation}
H_{\rm r-e}=\phi_i\sum_{q}g_{q}^i(a_{q,i}^{\dagger}+a_{q,i})\,,
\end{equation}
where  $g_{q}^i$ are the coupling strengths. Finally, the last term  describes the Josephson junction and it's capacitive coupling to the two oscillators:
\begin{equation}
H_{J}=-E_{J}\cos{\eta}-2e\bigg(V-\sum_{i=1,2}V_{\rm res}^i\bigg)N\,,
\end{equation} 
where  $E_J$, $\eta$, $V$, $N$, and $V_{\rm res}$ are the Josephson energy, the phase bias, the voltage bias, the number of Cooper pairs, and the voltage induced by the oscillators, respectively. The latter is given by  $V_{\rm res}=\sum_{i=1,2}V_{\rm res}^i$, with  
\begin{equation}
V_{\rm res}^i=-\left(\frac{\hbar}{2e}\dot{\phi}_i\right)\,;\,\,\,\dot{\phi}_i=\frac{i}{\hbar}[H_{\rm res}^i,\phi_i]\,.  
\end{equation}
Note that  we defined the  following conjugate variables for the cavities and the JJ, respectively: 
\begin{align}
[q_i,\phi_j]&=2ie\delta_{ij}\,,\\
[N,\eta]&=-i\,, 
\end{align} 
and we have that $E_J=E_{J,0}\cos{(\pi\Phi/\Phi_0)}$, with $E_{J,0}$ the Josephson energy and $\Phi$ the flux threaded through the JJ ($\Phi_{0}=h/2e$ is the flux quantum),  namely it can be controlled by controlling the flux $\Phi$ \cite{Hofheinz}. Next we exclude the Copper pairs number from the Hamiltonian by performing  a time-dependent unitary transformation on the full Hamiltonian, namely  \cite{Ankerhold2013}
\begin{equation}
\widetilde{H}_{\rm tot}=U_N(t)H_{\rm tot}U_N^{\dagger}(t)+i\,\partial_tU_N(t)U_N^{\dagger}(t)
\end{equation}
with
\begin{equation}
U_N(t)=e^{i(\omega_Jt+\phi_1+\phi_2)N}\,,
\end{equation}
where $\omega_J\equiv2eV/\hbar$. That affects both the resonators and JJ Hamiltonians so that they become:
\begin{align}
\widetilde{H}_{\rm res}^i&=\frac{q^2}{2C_i}+\left(\frac{\hbar}{2e}\right)\frac{\phi^2_i}{2L_i}\,\\
\widetilde{H}_{J}&=-E_J\cos{\bigg(\omega_Jt+\sum_{i=1,2}\phi_i\bigg)}\,,
\end{align} 
where $\tilde{q}_i=q_i+2eN$ is still continuos,although $N$ is discrete, so that the Hamiltonian of the resonators becomes identical to the original one. We can  quantize the excitations in the two resonators by writing 
\begin{equation}
\phi_i=\sqrt{\kappa_i}(a_i^\dagger-a_i),\,\,\,\,q_i=i\sqrt{\xi_i}(a_i^\dagger+a_i)
\end{equation}
with $\kappa_i=\hbar/2m_i\omega_i$,  $\xi_i=2m_i\omega_i\hbar$,  $m_i=(\hbar/2e^2)C_i$, and $a_i^\dagger$ ($a_i$) are the creation (annihilation) operators for the resulting photons (and which satisfy $[a_i,a_j^\dagger]=\delta_{ij}$).  Using this description, the oscillators Hamiltonian becomes:  
\begin{equation}
\widetilde{H}_{\rm res}^i=\omega_ia_i^\dagger a_i\,,
\end{equation}
with $\omega_i=1/\sqrt{L_iC_i}$. We note that the parameter $\kappa_i\equiv R_K^{-1}\sqrt{L_i/C_i}$ quantifies the range of the coupling regime: for $\kappa_i\ll1$ the setup is in the weak coupling limit which, combined with small photonic emission rate ($\propto E_J$), is well described by  the so called $P(E)$ theory  \cite{nazarov89,devoret90,girvin90}. For $\kappa_i>1$ instead, one can reach the strong coupling regime where non-linearities and  feed-back effects of the electronic transport on the cavity (and vice-versa) become manifest. In this paper we will only be concerned with the weak coupling limit, since most experiments are carried out in this regime.  

Until now, the description was exact, but the coupling between the JJ and the resonators is strongly non-linear. To make further progress, it is instructive to switch to the interaction picture with respect to the two cavities, which pertains to write:
\begin{equation}
\widetilde{H}_{J}(t)=-E_J\cos{\left[\omega_Jt+\sum_{i=1,2}\sqrt{\kappa_i}\left(a_i^\dagger e^{i\omega_it}+a_ie^{-i\omega_it}\right)\right]}\,,
\end{equation}
which is generally time-dependent. In the following we are interested in photonic processes that are resonant with $\omega_J$,  namely where the energy of the Cooper pairs gained from the bias voltage equals the energy of a given number of quanta in the two resonators. Here we are interested in the two-photon emission processes, one from each cavity,  such that the bias voltage is tuned to   $\omega_J=\omega_1+\omega_2$. 

Typically, one performs the so called rotating wave approximation (RWA), which amounts to keeping only the terms in the above Hamiltonian that do not depend explicitly on time. To identify such terms, we expand the argument of the cosine function so that we obtain: 
\begin{align}
\widetilde{H}_{J}(t)&=-\frac{E_J}{2}e^{-(\kappa_1+\kappa_2)/2}\left[\prod_{i=1,2}\sum_{n_i,m_i}e^{i\omega_Jt}\frac{i^{n_i+m_i}(\sqrt{\kappa_i})^{n_i+m_i}}{n_i!m_i!}\right.\nonumber\\
&\!\!\!\!\!\!\!\left.\times(a_i^\dagger)^{n_i}(a_i)^{m_i}e^{i(n_i-m_i)\omega_it}+{\rm h. c.}\right]\,.
\end{align}
Now we perform the RWA by keeping only the static terms, namely we impose the condition $n_1-m_1=n_2-m_2\equiv-1$. This statement is only valid if the two frequencies are incommensurate with $\omega_J$, namely $\omega_J/\omega_1\neq s/r$, with $s,r\in{\mathbf N}$ [and thus $\omega_J/\omega_2\neq r/(s-r)$].   Otherwise, we would have the condition
\begin{equation}
\frac{(n_1-m_1)s}{r}+\frac{(n_2-m_2)(r-s)}{r}=-1
\end{equation}
or
\begin{equation}
n_1-m_1=\left[(n_2-m_2)-(n_2-m_2+1)\frac{r}{s}\right]\,,
\end{equation} 
which implies that $(n_2-m_2+1)r/s=q\in{\mathbf N}$.  For example, if $r/s=1/2$, we have that $n_2-m_2+1=2q$ and $n_1-m_1=q-1$. We will not consider this commensurate limit any further here, and we leave it as a subject for future analysis. We mention that the single-photon and two-photon emission from a {\it single } cavity have been already discussed theoretically \cite{Lep2013,Ankerhold2013,Armour2013} and implemented experimentally \cite{Hofheinz}. Thus, even if the energy of these modes is the same, the resulting processes must not be similar to the single cavity case. We then obtain for the static part of the above Hamiltonian the following expression:
\begin{align}
H^{RWA}_{J}&=-\frac{\tilde{E}_J}{2}\left[\prod_{i=1,2}\sum_{n_i}\frac{(i\sqrt{\kappa_i})^{2n_i+1}}{n_i!(n_i+1)!}(a_i^\dagger)^{n_i}(a_i)^{n_i+1}+{\rm h. c.}\right]\nonumber\\
&\!\!\!=\frac{\tilde{E}_J}{2}:\left(a_1a_2+a_1^\dagger a_2^\dagger\right)\frac{J_1(2\sqrt{\kappa_1n_1})}{\sqrt{n_1}}\frac{J_1(2\sqrt{\kappa_2n_2})}{\sqrt{n_2}}:\,,
\end{align}
where $\tilde{E}_J=E_J\exp{[-(\kappa_1+\kappa_2)/2}]$, and $J_p(z)$ is the Bessel function of the first kind that has the form 
\begin{equation}
J_{p}(z)=\sum_n(-1)^n\frac{(z/2)^{2n+p}}{n!(n+p)!}\,,
\end{equation}
and $:\ldots:$ means normal ordering of the operators, i.e. all annihilation operators on the right and all the creation ones on the left.  We mention, however, that we still need to add the environments and their coupling to the two oscillators to complete the approximate setup. While we should in principle perform the same RWA on the external modes only (and on their coupling), it turns out that for $g^i_{q}\ll\omega_i$ we can safely utilize the initial coupling Hamiltonian \cite{carmichaelBook}.  In the following we analyze the photon emission by the junction by using both the density matrix approach and the input-output approach. We will calculate quantities such as the average photon number (or fluxes) and the second coherence factors that unravel the photonic statistics.

\section{Density Matrix Description}
\label{sec3}

In this section we analyze the photons in the cavity by resorting to the density matrix approach, similar to that used in Ref.~\onlinecite{Ankerhold2013}. Such a description allows us to analyze the photonic fields {\it inside} each of the two cavities.  Let us  start by writing down the equation of motion for the total photonic density matrix, including the external transmission lines (environments): 
\begin{equation}
\frac{d\rho_{tot}}{dt}=-\frac{i}{\hbar}[H_{J}^{RWA}+\sum_{i=1,2}(H_{\rm env}^i+H_{\rm r-e}^i),\rho_{tot}]\,.
\end{equation} 
so that the reduced density matrix $\rho_S(t)$ describing only the oscillators and the $JJ$ can be found by tracing the environment degrees of freedom, namely  $\rho_{S}(t)={\rm Tr}_{env}[\rho_{tot}(t)]$.  Within the Markov approximation,  and up to second order in the coupling $g^i_q$, one obtains the following (Lindblad) form for the system density matrix:  
\begin{equation}
\frac{d\rho_S}{dt}=\mathcal{L}\rho_S=-\frac{i}{\hbar}[H_{J}^{RWA},\rho_S]+\sum_{i=1,2}\gamma_i\left[2a_i\rho_Sa_i^\dagger-\{a_i^\dagger a_i,\rho_S\}\right]\nonumber\,,
\end{equation}
which describes the dynamics of the electromagnetic modes in the cavity, where $\gamma_i=2\pi\sum_{q}|g_{q}^i|^2$ is the $i$th  cavity decay rate. Note that the expectation values of an operator $\hat{X}$ reads $\langle \hat{X}(t)\rangle={\rm Tr}_{sys}[\hat{X}\rho_S(t)]$, where the trace is taken over the oscillators degrees of freedom. In order to characterize the photonic emission  statistics, we define the following second-order correlation function \cite{MilburnBook}:
\begin{equation}
G^{(2)}_{pj}(t,\tau)=\langle:n_p(t)n_j(t+\tau):\rangle=\langle:n_j(0)e^{\mathcal{L}\tau}[\rho_S(t)n_p(0)]:\rangle\,,
\end{equation}
which describe the probability to detect a photon in branch $p$ at time $t$ and one photon in branch $j$ at time $t+\tau$,  with  $\tau$ being the time delay between the photonic counts.  For $p\neq j$ ($p=j$) they correspond the cross-correlation (auto-correlation) of the photonic emission, namely detection of photons from different (the same) emitter. In the stationary limit, for $t\rightarrow\infty$, the correlators depend only on the time delay $\tau$ \cite{MilburnBook}.  One defines a normalized second correlation function (for $t\rightarrow\infty$) as
\begin{equation}
g_{pj}^{(2)}(t, \tau)=\frac{G^{(2)}_{pj}(t,\tau)}{\langle n_p(t)\rangle\langle n_j(t+\tau)\rangle}\,,
\end{equation}
which, in the stationary limit and for $\tau=0$ (zero time delay) becomes:
\begin{equation}
g_{pj}^{(2)}(0)=1+\frac{V(n_p,n_j)-\delta_{pj}\langle n_j\rangle}{\langle n_p\rangle\langle n_j\rangle}\,,
\end{equation}
with $V(n_p,n_j)=\langle n_pn_j\rangle-\langle n_p\rangle\langle n_j\rangle$ being the variance of the field.  By accessing  the $g_{pj}^{(2)}(0)$ one can infer the degree of ``quantumness" of the emitted light. Let us now analyze in detail this statement, by comparing the possible outcomes from both the classical and quantum descriptions.  The correlation between the photon emission processes from the two oscillators is encoded in the variance ${\rm Var}(n_1-n_2)=\langle (\delta n_1-\delta n_2)^2\rangle\geq0$.  Assuming $\langle n_1\rangle=\langle n_2\rangle=\langle n\rangle$ (equally populated cavities), this implies that $\langle \delta n_1^2\rangle+\langle \delta n_2^2\rangle\geq 2\langle \delta n_1\delta n_2\rangle$. For a {\it classical}  field, we can define  $g_{c,pj}^{(2)}(0)=\langle \delta n_p\delta n_j\rangle/n^2$, {\it without} normal ordering since $n_i$  are not operators, but random variables. By using the above condition, we  obtain for {\it classical} states   the following  Cauchy-Schwarz inequalities:
\begin{align}
g_{c,12}^{(2)}(0)\leq\frac{g_{c,11}^{(2)}(0)+g_{c,22}^{(2)}(0)}{2},\,\,\,g_{c,12}^{(2)}(0)\leq\sqrt{g_{c,11}^{(2)}(0)g_{c,22}^{(2)}(0)}\,.
\label{CS}
\end{align}
For a {\it quantum} field instead, $\langle \delta n_p\delta n_j\rangle=\langle:\delta n_p\delta n_j:\rangle+\delta_{pj}\langle n_p\rangle$ and,  since in this case $g_{q,pj}^{(2)}(0)=\langle:\delta n_p\delta n_j:\rangle$, we obtain that $\langle \delta n_p\delta n_j\rangle/\langle n\rangle^2=g_{q,pj}^{(2)}(0)+(1/\langle n\rangle)\delta_{pj}$. Thus, for a quantum field we obtain the less stringent condition:
\begin{equation}
g_{q,12}^{(2)}(0)\leq\frac{g_{q,11}^{(2)}(0)+g_{q,22}^{(2)}(0)}{2}+\frac{1}{n}\,,
\end{equation}
which implies that the  Cauchy-Schwarz inequalities can be violated. 
 
Next we analyze explicitly these Cauchy-Schwarz inequalities for our two-photon setup.   For that, we first calculate  various average quantities in the stationary limit $d\langle \hat{X}\rangle/dt=0$,  so that  for the average photon number $\hat{X}=\hat{n}_j$ and photon number product $\hat{X}=\hat{n}_p\hat{n}_j$,  we obtain the following relations:
\begin{align}
\langle n_p\rangle&=-\frac{i}{2\gamma_j}\langle[n_p,H_J^{RWA}]\rangle\,,\\
\langle n_pn_j\rangle=&-\frac{i}{2(\gamma_j+\gamma_p)}\langle[n_pn_j,H_J^{RWA}]\rangle+\frac{1}{2}\langle n_j\rangle\delta_{pj}\,.
\end{align}
Because the Hamiltonian $H_J^{RWA}$ is not quadratic in the field operators, the system of equations for  does not close, and we find:
\begin{align}
[n_j,H_J^{RWA}]&=\tilde{E}_J\sqrt{\kappa_j}:\left(a_{\bar{j}}a_j-a_{\bar{j}}^\dagger a_j^\dagger\right)\nonumber\\
&\!\!\!\!\!\!\times\left[J_0(2\sqrt{\kappa_jn_j})+J_2(2\sqrt{\kappa_jn_j})\right]\frac{J_1(2\sqrt{\kappa_{\bar{j}}n_{\bar{j}}})}{\sqrt{n_{\bar{j}}}}:\,,
\end{align}
and a similar, but more complicated result for the $[n_pn_j,H_J^{RWA}]$ commutator (not shown). To make progress, in the following we recall again the weak coupling assumption,  $\kappa_{1,2}\ll1$, as in the experiments carried out in Ref.~\onlinecite{Hofheinz}, and consider also that the two cavities are populated on average by only a few photons, so that   $\kappa_i\langle n_i\rangle\ll1$.  The latter condition depend on the parameter range and has to be checked self-consistently at the end of the calculation. Under these assumptions, we can approximate the system Hamiltonian as follows:
\begin{align}
H_J^{RWA}&\approx \tilde{E}_J\sqrt{\kappa_1\kappa_2}:(a_1^\dagger a_2^\dagger+a_1a_2):\,,
\end{align}
neglecting terms of the order $\mathcal{O}(\kappa_in_i)^2$. This simple quadratic Hamiltonian describes the well known non-degenerate parametric amplifier \cite{MilburnBook}. For the photon number expectation values in the two cavities  we get:
\begin{equation}
\langle n_j\rangle=\frac{\beta^2}{2}\left(\langle n_j\rangle+\langle n_p\rangle+1\right)\,,
\end{equation}
where $\beta=(2\tilde{E}_J\sqrt{\kappa_1\kappa_2}/\gamma)$  for the case of two identical cavities ($\gamma_1=\gamma_2$), so that $\langle n_1\rangle=\langle n_2\rangle\equiv\langle n\rangle$, with
\begin{equation}
\langle n\rangle=\frac{\beta^2}{2(1-\beta^2)}\,.
\end{equation}
We see that $\beta=1$  is the threshold for an instability at which the cavities become coherently populated instead of incoherently, as it happens below this threshold. We will not discuss any further the behavior above this threshold, as we are interested here by the incoherent regime.  However, note that for the quadratic approximation to be valid, the condition $\kappa_i\langle n_i\rangle\ll$ must be met, and thus the threshold cannot be achieved within this limit as strong quantum fluctuation may change the critical condition \cite{Padurariu2012}.  
Under all these assumptions and after some lengthy, but straightforward calculations we obtain: 
\begin{align}
\langle n_jn_p\rangle&=\frac{\beta^2[\langle n^2\rangle+(1+\beta^2)\langle n\rangle+2\langle n_1n_2\rangle+\beta^2/2]}{4-\beta^2}\nonumber\\
&+\frac{\beta^2}{4}(2\langle n\rangle+1)+\frac{1}{2}\langle n\rangle\delta_{jp}\,,
\end{align}
or 
\begin{align}
\langle n^2\rangle&=A\langle n^2\rangle+B\langle n_1n_2\rangle+C\,,\\
\langle n_1n_2\rangle&=A\langle n^2\rangle+B\langle n_1n_2\rangle+D\,,
\end{align}
with
\begin{align}
A&=\frac{\beta^2}{4-\beta^2},\nonumber\\
C&=\frac{\beta^4}{(1-\beta^2)(4-\beta^2)}+\frac{\beta^2}{2(1-\beta^2)}\,,\\
D&=\frac{\beta^4}{(1-\beta^2)(4-\beta^2)}+\frac{\beta^2}{4(1-\beta^2)}\,.
\end{align}
and $B=2A$. By further manipulating the above expression, we finally obtain:
\begin{align}
\langle n^2\rangle&=2(\langle n\rangle)^2+\langle n\rangle\,,\\
\langle n_1n_2\rangle&=\frac{\langle n\rangle(1+4\langle n\rangle)}{2}\,.
\end{align}
which in turn leads to
\begin{align}
g^{(2)}_{11}(0)&=g^{(2)}_{22}(0)=2,\nonumber\\
g^{(2)}_{12}(0)&=2+\frac{1}{2\langle n\rangle}\,.
\end{align}
We mention that for $g_{ii}^{(2)}(0)=2$ the radiation resembles the thermal radiation, although the system is at $T=0$, and it is characteristic to the parametric non-degenerate oscillator. The cross-correlation instead, $g_{12}^{(2)}(0)\sim1/2\langle n\rangle$ for $\langle n\rangle\ll1$, meaning strong bunching of the emitted radiation at low emission rates $\propto\langle n\rangle$.     
In order to check the non-classical character of the emitted pairs, we calculate the so called noise reduction factor ($NRF$) given by
\begin{equation}
NRF=\frac{{\rm Var}(n_1-n_2)}{2\langle n\rangle}=\frac{\langle n\rangle}{2}\left(g_{11}^{(2)}+g_{22}^{(2)}-2g_{12}^{(2)}\right)+1\,,
\end{equation}
which is $NRF\geq1$ for classical light, when the photon emission is uncorrelated, and can be  $0<NRF<1$ when quantum correlations are manifest.  For the few photon limit considered here, we find $NRF=1/2$, which means the photon pair emission shows strong quantum correlations, which should be easily detected in experiments. The very same relation implies that the  classical Cauchy-Schwarz inequalities in Eq.~\ref{CS} are violated. 

We mention that for larger emission rates ($\propto E_J$) or larger $\kappa_i$'s, one can go beyond the few photon limit and access higher non-linearities of the $JJ$. Such a regime, although interesting, is left for a future study \cite{TrifSimonUnpublished}.

\section{Input-output description}
\label{sec4}

While the density matrix approach allows us to calculate the field in the cavity, it does not tell us the properties of the field exiting in the transmission lines and measured by the detectors. In order to address this issue, we resort to the input-output description of the system: an input field is sent to the combined system formed by the  JJ and the two oscillators, and the output field is measured. The input field could even be just the quantum vacuum, as it will be assumed in the following.     

In Fig.~\ref{sketch_JJ} we show a schematics of the input-output fields. We assume that each oscillator is coupled to its  external transmission line (environment) by only one side, so that there is only one input and output fields, respectively.  The  relations between the input, output, and the cavity fields for oscillator $\alpha=1,2$ are as follows \cite{MilburnBook}: 
\begin{align}
\label{eq_motion_1}
\!\!\!\!\!\dot{a}_\alpha(t)=&-\frac{i}{\hbar}[a_{\alpha}(t),H_J^{RWA}(t)]-\frac{\gamma_\alpha}{2}a_\alpha(t)-\sqrt{\gamma_\alpha}b_{in,\alpha}(t)\,,\\
b_{out,\alpha}(t)&=b_{in,\alpha}(t)+\sqrt{\gamma_\alpha}a_\alpha(t)\,.
\label{eq_motion}
\end{align}
where $b_{in}^{\alpha}(t)$ [$b_{out}^{\alpha}(t)$] are the input (output) fields defined as follows:
\begin{align}
b_{in,\alpha}(t)&=\frac{1}{\sqrt{2\pi \rho_\alpha}}\sum_{q}e^{-i\omega_{q,\alpha}(t-t_0)}a_{q,\alpha}(t_0)\,,\\
b_{out,\alpha}(t)&=\frac{1}{\sqrt{2\pi \rho_\alpha}}\sum_{q}e^{-i\omega_{q,\alpha}(t-t_1)}a_{q,\alpha}(t_1)\,,
\label{in_and_out}
\end{align}
with $t_0<t<t_1$, and $\rho_\alpha=\sum_q\delta(\omega_\alpha-\omega_{q,\alpha})$ is the environment $\alpha$ density of states. 
We mention that for deriving these expression we assumed the RWA to hold, i.e. we neglected the  terms of the form $a_{\alpha}^\dagger a_{q,\alpha}^\dagger$ and $a_{\alpha} a_{q,\alpha}$ (counter rotating terms). Because of the second term in Eq.~\eqref{eq_motion_1} (the commutator), the equation of motion is highly non-linear, and  reads: 
\begin{align}
&[a_{\alpha},H_J^{RWA}(t)]=\tilde{E}_J\sqrt{\kappa_\alpha}\nonumber\\
&\times:\left[a_{\bar{\alpha}}\frac{a_\alpha}{a_\alpha^\dagger}J_2(2\sqrt{\kappa_\alpha n_\alpha})-a_{\bar{\alpha}}^\dagger J_0(2\sqrt{\kappa_\alpha n_\alpha})\right]\frac{J_1(2\sqrt{\kappa_{\bar{\alpha}}n_{\bar{\alpha}}})}{\sqrt{n_{\bar{\alpha}}}}:\nonumber\\
&\!\!\!\!\!\approx-\tilde{E}_J\sqrt{\kappa_\alpha\kappa_{\bar{\alpha}}}a_{\bar{\alpha}}^\dagger+\mathcal{O}[(\kappa_\alpha n_{\alpha})^2]\,,
\end{align}
where in the last line we presuppose the limit  $\kappa_\alpha n_\alpha\ll1$. Moreover, we assume the two oscillators have large $Q$-factors so that their overlap in frequency is negligible, i.e. $|\omega_{1}-\omega_2|\gg\gamma_{1,2}$.  

Utilizing the same quadratic approximation of the JJ Hamiltonian, we arrive at the following equation of motion for the cavity field operators:
\begin{align}
\dot{a}_{\alpha}&=-i\omega_\alpha a_\alpha(t)-i\tilde{E}_J\sqrt{\kappa_\alpha}a^\dagger_{\bar{\alpha}}(t)e^{-i(\omega_1+\omega_2)t}\nonumber\\
&-\frac{\gamma_\alpha}{2}a_\alpha(t)-\sqrt{\gamma_\alpha}b_{in, \alpha}(t)\,,
\end{align}
which implies that the left and right fields are coupled and we need to solve the dynamics for both of them at the same time. It is instructive to switch to the Fourier space in order to obtain the needed relationships
\begin{equation}
a_{\alpha}(t)=\frac{1}{\sqrt{2\pi}}\int_{-\infty}^\infty d\omega e^{i\omega t}a_\alpha(\omega)\,,
\end{equation}
and the commutation relations:
\begin{equation}
[a^{\dagger}_\alpha(\omega),a_\alpha(\omega')]=\delta(\omega-\omega')\,,
\end{equation}
and similarly for the other operators. This in turn allows us to write the following equation relating the input and cavity fields:
\begin{align}
\left[i(\omega-\omega_\alpha)-\frac{\gamma_\alpha}{2}\right]a_\alpha(\omega)&+\tilde{E}_J\sqrt{\kappa_\alpha}a^\dagger_{\bar{\alpha}}(\omega_1+\omega_2-\omega)\nonumber\\
&=\sqrt{\gamma_\alpha}b_{in,\alpha}(\omega)\,.
%\left[i(\omega-\omega_{\bar{\alpha}})+\frac{\gamma_{\bar{\alpha}}}{2}\right]a^\dagger_{\bar{\alpha}}(\omega)&+i\tilde{E}_Ja_{\alpha}(\omega_1+\omega_2-\omega)=-\sqrt{\gamma_{\bar{\alpha}}}b^\dagger_{in,\bar{\alpha}}(\omega)\rightarrow\nonumber\\
%a^\dagger_{\bar{\alpha}}(\omega_1+\omega_2-\omega)=\frac{\sqrt{\gamma_{\bar{\alpha}}}}{i(\omega-\omega_\alpha)-\gamma_{\bar{\alpha}}/2}b^\dagger_{in, \bar{\alpha}}(\omega_1+\omega_2-\omega)&+\frac{i\tilde{E}_J}{i(\omega-\omega_\alpha)-\gamma_{\bar{\alpha}}/2}a_{\alpha}(\omega)\rightarrow\nonumber\\
%\left[-i(\omega-\omega_\alpha)+\frac{\gamma_\alpha}{2}+\frac{\tilde{E}_J^2}{i(\omega-\omega_\alpha)-\gamma_{\bar{\alpha}}/2}\right]a_\alpha(\omega)&=-\sqrt{\gamma_\alpha}b_{in,\alpha}(\omega)+\frac{i\tilde{E}_J\sqrt{\gamma_{\bar{\alpha}}}}{i(\omega-\omega_\alpha)-\gamma_{\bar{\alpha}}/2}b^\dagger_{in, \bar{\alpha}}(\omega_1+\omega_2-\omega)\,
\end{align}
Manipulating this relation and its hermitian conjugate, and utilizing  the relation between the input, output and cavity fields in Eq.~\eqref{in_and_out}, we obtain:
%\begin{align}
%a_{\alpha}(\omega)&=\frac{1}{-i(\omega-\omega_\alpha)+\frac{\gamma_\alpha}{2}+\frac{\tilde{E}_J^2}{i(\omega-\omega_\alpha)-\gamma_{\bar{\alpha}}/2}}\nonumber\\
%&\times\left[-\sqrt{\gamma_\alpha}b_{in,\alpha}(\omega)+\frac{i\tilde{E}_J\sqrt{\gamma_{\bar{\alpha}}}}{i(\omega-\omega_\alpha)-\gamma_{\bar{\alpha}}/2}b^\dagger_{in, \bar{\alpha}}(\omega_1+\omega_2-\omega)\right]
%\end{align}
%Let us rewrite this expression in a more compact form:
\begin{align}
a_\alpha(\omega)&=A_\alpha(\omega)b_{in,\alpha}(\omega)+B_\alpha(\omega)b_{in,\bar{\alpha}}^\dagger(\omega_1+\omega_2-\omega)\,,\\
b_{out,\alpha}(\omega)&=[\sqrt{\gamma_\alpha}A_\alpha(\omega)+1]b_{in,\alpha}(\omega)\nonumber\\
&+\sqrt{\gamma_{\alpha}}B_\alpha(\omega)b^\dagger_{in,\bar{\alpha}}(\omega_1+\omega_2-\omega)\,,
\end{align}
with 
\begin{align}
A_\alpha(\omega)&=-\frac{\sqrt{\gamma_\alpha}}{-i(\omega-\omega_\alpha)+\frac{\gamma_\alpha}{2}+\frac{\tilde{E}_J^2\kappa_\alpha\kappa_{\bar{\alpha}}}{i(\omega-\omega_\alpha)-\gamma_{\bar{\alpha}}/2}}\,,\\
B_\alpha(\omega)&=\frac{i\tilde{E}_J\sqrt{\kappa_\alpha\kappa_{\bar{\alpha}}\gamma_{\bar{\alpha}}}}{\left[i(\omega-\omega_\alpha)-\frac{\gamma_{\bar{\alpha}}}{2}\right]\left[-i(\omega-\omega_\alpha)+\frac{\gamma_\alpha}{2}+\frac{\tilde{E}_J^2\kappa_\alpha\kappa_{\bar{\alpha}}}{i(\omega-\omega_\alpha)-\gamma_{\bar{\alpha}}/2}\right]}\,.
\end{align}
%At the resonances, we obtain the simplified expressions:
%\begin{align}
%\!\!\!A_\alpha(\omega_\alpha)&=-\frac{\sqrt{\gamma_\alpha}}{\frac{\gamma_\alpha}{2}-\frac{2\tilde{E}_J^2\kappa_{\alpha}\kappa_{\bar{\alpha}}}{\gamma_{\bar{\alpha}}}},\,B_\alpha(\omega_\alpha)=-\frac{2i\tilde{E}_J}{\sqrt{\gamma_{\bar{\alpha}}}\left(\frac{\gamma_\alpha}{2}-\frac{2\tilde{E}_J^2\kappa_{\alpha}\kappa_{\bar{\alpha}}}{\gamma_{\bar{\alpha}}}\right)}\,.
%\end{align}
We are now in position to use the above findings to  calculate the relevant observables and correlators. However, we will analyze both the case where the detectors have infinite bandwidth, namely all photons emitted  are collected, as well as the finite bandwidth case, and state the differences compared to the density matrix approach.   Let us start with the infinite bandwidth case, and then discuss briefly the implications of  finite bandwidth detection. 

\subsection{Infinite bandwidth detection}

Here we assume that the efficiency of the detectors is unity and that their bandwidth is infinite, thus they are collecting all the emitted photons.  The outgoing photonic flux from oscillator $\alpha$ reads: 
\begin{align}
\Gamma_{\alpha}(t)&=\langle b_{out,\alpha}^\dagger(t)b_{out,\alpha}(t)\rangle=\frac{1}{2\pi}\int_{-\infty}^\infty d\omega\int_{-\infty}^\infty d\omega'e^{i(\omega-\omega')t}\nonumber\\
&\!\!\!\!\!\!\!\!\!\times\langle b^\dagger_{out,\alpha}(\omega)b_{out,\alpha}(\omega')\rangle=\frac{\gamma_\alpha}{\sqrt{2\pi}}\int_{-\infty}^{\infty}d\omega|B_\alpha(\omega)|^2=\frac{\sqrt{2\pi}\beta^2\gamma_\alpha}{2(1-\beta^2)}\,,
\label{Rate_Gamma}
\end{align}  
which, by assuming  $\gamma_{1}=\gamma_2\equiv\gamma$ implies $\Gamma_\alpha\equiv\Gamma$, while the autocorrelation and cross-correlation functions at $\tau=0$ respectively read:
\begin{align}
G_{\alpha\alpha}^{(2)}(0)&=\langle b^\dagger_{out,\alpha}(t)b^\dagger_{out,\alpha}(t) b_{out,\alpha}(t)b_{out,\alpha}(t)\rangle\nonumber\\
&=\frac{\gamma_\alpha^2}{\pi}\left[\int_{-\infty}^\infty d\omega|B_{\alpha}(\omega)|^2\right]^2=\frac{\pi\beta^4}{(1-\beta^2)^2}\,,\\
G_{\alpha\bar{\alpha}}^{(2)}(0)&=\langle b^\dagger_{out,\alpha}(t)b^\dagger_{out,\bar{\alpha}}(t) b_{out,\bar{\alpha}}(t)b_{out,\alpha}(t)\rangle\nonumber\\
&\!\!\!\!\!\!\!\!\!\!=\frac{1}{2\pi}\int d\omega\int d\omega'\big[B^*_\alpha(\omega)A^*_{\bar{\alpha}}(\omega_1+\omega_2-\omega)\hspace{-1cm}\nonumber\\
&\times B_{\bar{\alpha}}(\omega')A_\alpha(\omega_1+\omega_2-\omega')+|B_\alpha(\omega)|^2|B_{\bar{\alpha}}(\omega')|^2\big]\nonumber\\
&=\frac{\pi\gamma^2\beta^2(1+\beta^2)}{2(1-\beta^2)^2}\,.
\end{align}
From these expression, we can readily evaluate the zero time delay ($\tau=0$) second-order coherence functions:
\begin{align}
g^{(2)}_{\alpha\alpha'}(0)&=\frac{\langle b^\dagger_{out,\alpha}(t)b^\dagger_{out,\alpha}(t) b_{out,\alpha}(t)b_{out,\alpha}(t)\rangle}{\langle b^\dagger_{out,\alpha}(t)b_{out,\alpha}(t)\rangle\langle b^\dagger_{out,\alpha}(t)b_{out,\alpha}(t)\rangle}\nonumber\\
&=\left\{
\begin{array}{cc}
2, & \alpha'=\alpha\\
2+{\displaystyle \frac{\sqrt{\pi}\gamma}{\sqrt{2}\Gamma}}, & \alpha'=\bar{\alpha}\,.
\end{array}
\right.
\end{align} 
We see that in the infinite bandwidth limit we obtain again the same result we found from the density matrix approach for the autocorrelation function, $g_{\alpha\alpha}^{(2)}=2$,  while $g_{12}^{(2)}(0)\propto1/\Gamma$ for $\Gamma/\gamma\ll1$.  Note that experimentally it is measured not the cavity field, but the external photon flux and  the above results are due to the well-known relationships  
\begin{align}
\Gamma_{\alpha}=\langle b^\dagger_{out,\alpha}(t)b_{out,\alpha}(t)\rangle&=\gamma_\alpha\langle a^\dagger_\alpha(t)a_\alpha(t)\rangle\,,\\
\langle b^\dagger_{out,\alpha}(t)b_{out,\alpha}^\dagger(t) b_{out,\alpha}(t)b_{out,\alpha}(t)\rangle&=\gamma_\alpha^2\langle a^\dagger_\alpha(t)a^\dagger_\alpha(t) a_\alpha(t)a_\alpha(t)\rangle\,.
\label{correspondence}
\end{align}
which hold in the infinite bandwidth case.  Once again, deviations from the quadratic Hamiltonian lead to changes in the second order coherence factors, of the order $\mathcal(\kappa\Gamma/\gamma)$. Since we consider the few photon regime, such corrections are negligible. It worth mentioning that the due to the correspondence in Eq.~\eqref{correspondence} the Cauchy-Schwarz inequalities are in the same way as described in the previous section, thus the emitted light from the cavities is non-classical.    

Next we calculate the time-dependence of the $g^{(2)}_{\alpha\alpha'}(\tau)$ function in order to reveal the dynamics of the two resonators, and which is defined as follows:
\begin{equation}
g_{\alpha\alpha'}^{(2)}(\tau)=\frac{G_{\alpha\alpha'}^{(2)}(\tau)}{\Gamma^2}\,.
\label{finite_time}
\end{equation}
 The auto-correlation and cross-correlation second order coherence functions are given respectively by:
\begin{align}
G_{\alpha\alpha}^{(2)}(\tau)&=\langle b^\dagger_{out,\alpha}(t+\tau)b^\dagger_{out,\alpha}(t) b_{out,\alpha}(t)b_{out,\alpha}(t+\tau)\rangle\nonumber\\
&\!\!\!\!\!\!\!\!\!\!=\frac{\pi\gamma^2\beta^4}{2(1-\beta^2)^2}\left[1+e^{-\gamma\tau}\left(\cosh(\tilde{E}_J\tau)+\beta\sinh(\tilde{E}_J\tau)\right)^2\right]\nonumber\,,\\
G_{\alpha\bar{\alpha}}^{(2)}(\tau)&=\langle b^\dagger_{out,\alpha}(t+\tau)b^\dagger_{out, \bar{\alpha}}(t) b_{out,\bar{\alpha}}(t)b_{out,\alpha}(t+\tau)\rangle\nonumber\\
&=\frac{\pi\gamma^2\beta^4}{2(1-\beta^2)^2}\left[1+e^{-\gamma\tau}\left(\beta\cosh(\tilde{E}_J\tau)+\sinh(\tilde{E}_J\tau)\right)^2\right]\,,
\end{align}
which leads to the following normalized coherence functions:
\begin{align}
g_{\alpha\alpha'}^{(2)}(\tau)=1+\left\{
\begin{array}{cc}
\displaystyle{\frac{e^{-\gamma|\tau|}\left[\beta\cosh(\beta\gamma|\tau|)+\sinh(\beta\gamma|\tau|)\right]^2}{\beta^2}}\,, & \alpha'=\alpha\,\\
\displaystyle{\frac{e^{-\gamma|\tau|}\left[\cosh(\beta\gamma|\tau|)+\beta\sinh(\beta\gamma|\tau|)\right]^2}{\beta^2}}\,, & \alpha'=\bar{\alpha}\,.
\end{array}
\right.
\end{align}
These functions are witnesses of the dynamics of of oscillaltors as they  reveal their effective linewidth and thus their trend towards the instability point $\beta=1$.    Having found these functions, we can directly relate them to the  non-symmetrized photonic frequency noise, defined as follows:
\begin{align}
S_{\alpha\alpha'}(\omega)&=\int_{-\infty}^\infty d\tau\langle\hat{\Gamma}_{\alpha}(\tau)\hat{\Gamma}_{\alpha'}(0)\rangle e^{-i\omega\tau}\nonumber\\
&\equiv\Gamma^2\int_{-\infty}^\infty d\tau\left[g_{\alpha\alpha'}^{(2)}(\tau)-1\right]e^{-i\omega\tau}\,,
\end{align}  
with $\hat{\Gamma}_{\alpha}(\tau)=b^\dagger_{out,\alpha}(\tau)b_{out,\alpha}(\tau)$ representing the photonic flux operator out of the oscillator $\alpha$. That allows us to extract the noise to signal ratio, or the  Fano factor $F_{\alpha\alpha'}=S_{\alpha\alpha'}(\omega)/\Gamma$. This offers an estimate for  the number of photons that are correlated in the emission process \cite{Padurariu2012}, and it is given by:
\begin{equation}
F_{\alpha\alpha'}=\Gamma\int_{-\infty}^\infty d\tau\left[g_{\alpha\alpha'}^{(2)}(\tau)-1\right]\,.
\end{equation}
Using the expression for the photonic rate $\Gamma$ in Eq.~\eqref{Rate_Gamma}, and expressing the coefficient $\beta$ in terms of this quantity, we obtain for the auto-correlation Fano factor:
\begin{align}
F_{\alpha\alpha'}=\left\{
\begin{array}{cc}
\displaystyle{\frac{5\Gamma}{\gamma}+\frac{4\Gamma^2}{\gamma^2}}, & \alpha'=\alpha\\
\displaystyle{2+\frac{5\Gamma}{\gamma}+\frac{4\Gamma^2}{\gamma^2}}, & \alpha'=\bar{\alpha}'\,.
\end{array}
\right. 
\end{align}
We see that for $\Gamma/\gamma\gg1$ (but so that $\kappa\Gamma/\gamma<1$), $F\propto\Gamma^2$, which is a signature for strong photon bunching.  We note that for a Poisson process $F=1$, while for a thermal distribution $F=2$.   

\begin{figure}[t]
\begin{center}
\includegraphics[width=0.99\linewidth]{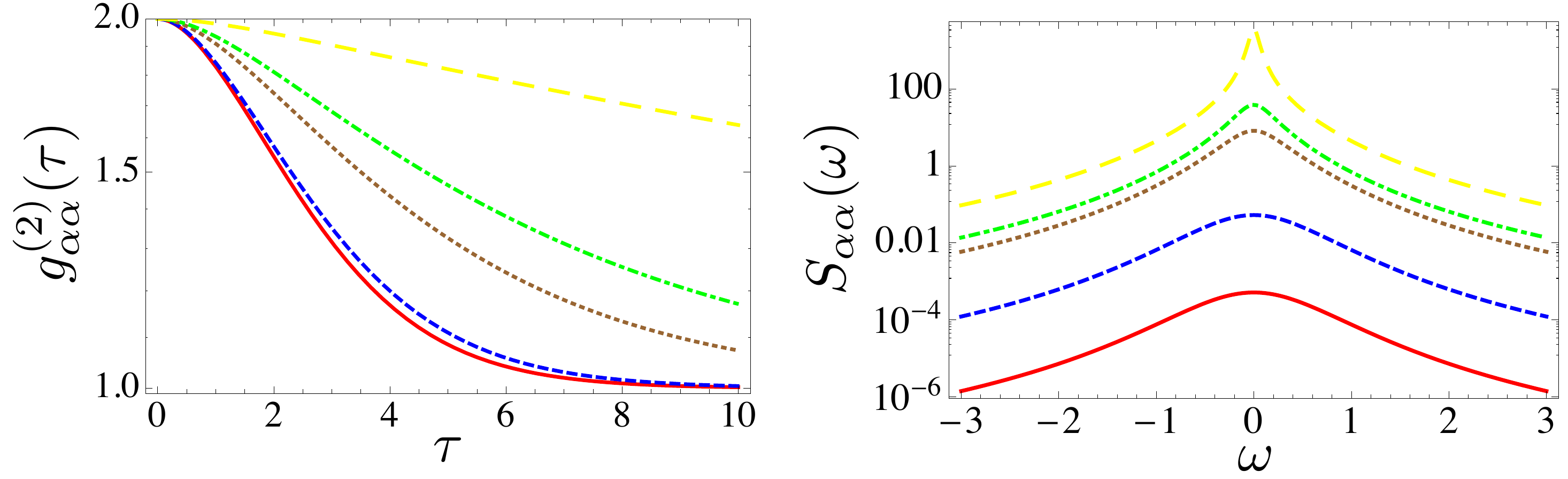}
\caption{Left plot: The time  dependence of the second-order auto-correlation function $g_{\alpha\alpha}^{(2)}(\tau)$  for different values of $\beta\propto\tilde{E}_J$. Right plot: The frequency  dependence of the auto-correlation noise  function $S_{\alpha\alpha}(\tau)$  for different values of $\beta\propto\tilde{E}_J$. In both plots,  the red, blue, brown, green, and yellow curves corresponds to $\beta=0.1, 0.3, 0.7, 0.8$, and $0.95$, respectively. Here the time scales are all expressed in terms of $\gamma$, and we recall that $\Gamma\propto\beta^2$.}
\label{auto_corr_inf}
\end{center}
\end{figure}

\begin{figure}[t]
\begin{center}
\includegraphics[width=0.99\linewidth]{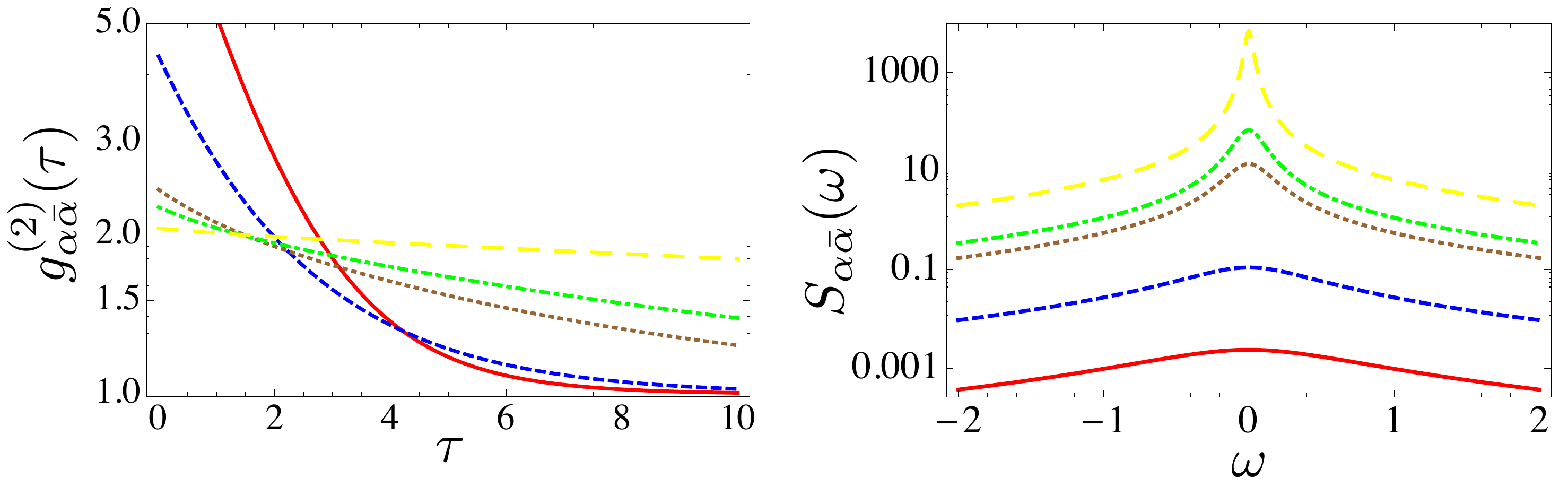}
\caption{Left plot: The time  dependence of the second-order cross-correlation function $g_{\alpha\bar{\alpha}}^{(2)}(\tau)$  for different values of $\beta\propto\tilde{E}_J$. Right plot: The frequency  dependence of the auto-correlation noise  function $S_{\alpha\bar{\alpha}}(\tau)$  for different values of $\beta\propto\tilde{E}_J$. In both plots,  the red, blue, brown, green, and yellow curves corresponds to $\beta=0.1, 0.3, 0.7, 0.8$, and $0.95$, respectively.}
\label{cross_corr_inf}
\end{center}
\end{figure}

In Fig.~\ref{auto_corr_inf} we plot the dependence of the functions $g_{\alpha\alpha}^{(2)}(\tau)$ on the time delay (left) and the autocorrelated noise $S_{\alpha\alpha}(\omega)$ (right). We see that for $\tau\rightarrow\infty$ $g^{(2)}_{\alpha\alpha}\rightarrow1$, while from the right plot we see that the linewidth becomes narrower as the emission rate $\propto E_J$ increases to values close to the instability threshold.  In Fig.~\ref{cross_corr_inf} on the other hand, we plot the dependence of the functions $g_{\alpha\bar{\alpha}}^{(2)}(\tau)$ on the time delay (left) and the cross-correlated noise $S_{\alpha\bar{\alpha}}(\omega)$ (right), which shows a similar behavior.

\subsection{Finite bandwidth detection}

In this section we discuss the properties of the outgoing light when the detection is taken within a finite frequency window $\Delta\omega$ around both  $\omega_1$ and $\omega_2$. For simplicity, we assume the same frequency  window for both oscillators.  We already expect that if $\Delta\omega\gg\gamma$ the results to be practically the same as in the previous section. However, when $\Delta\omega\leq\gamma$, one expects the photon statistics to be affected,  and we aim at quantifying such changes.  

The photon flux detected within the bandwidth $\Delta\omega$ reads:
\begin{align}
\Gamma_{r,\alpha}(t)&=\langle b_{out,\alpha}^\dagger(t)b_{out,\alpha}(t)\rangle_{\Delta\omega}=\frac{\gamma}{\sqrt{2\pi}}\int_{\omega_\alpha-\Delta\omega/2}^{\omega_\alpha+\Delta\omega/2}d\omega|B_\alpha(\omega)|^2\nonumber\\
&=\frac{\gamma\beta}{\sqrt{2\pi}}\left[\frac{\arctan\left(\frac{r}{1-\beta}\right)}{1-\beta}-\frac{\arctan\left(\frac{r}{1+\beta}\right)}{1+\beta}\right]\,,
\label{Rate_Gamma_FB}
\end{align}
where $r=\Delta\omega/\gamma$. The rate becomes Eq.~\eqref{Rate_Gamma} for $\Delta\omega\gg\gamma$ up to corrections $(\gamma/\Delta\omega)^3$. For $r<1$, we obtain:
\begin{equation}
\Gamma_{\alpha}^{\Delta\omega}\approx\frac{4\beta^2\Delta\omega}{(1-\beta^2)^2}+\mathcal{O}[r^2]\,.
\end{equation}

\begin{figure}[t]
\begin{center}
\includegraphics[width=0.99\linewidth]{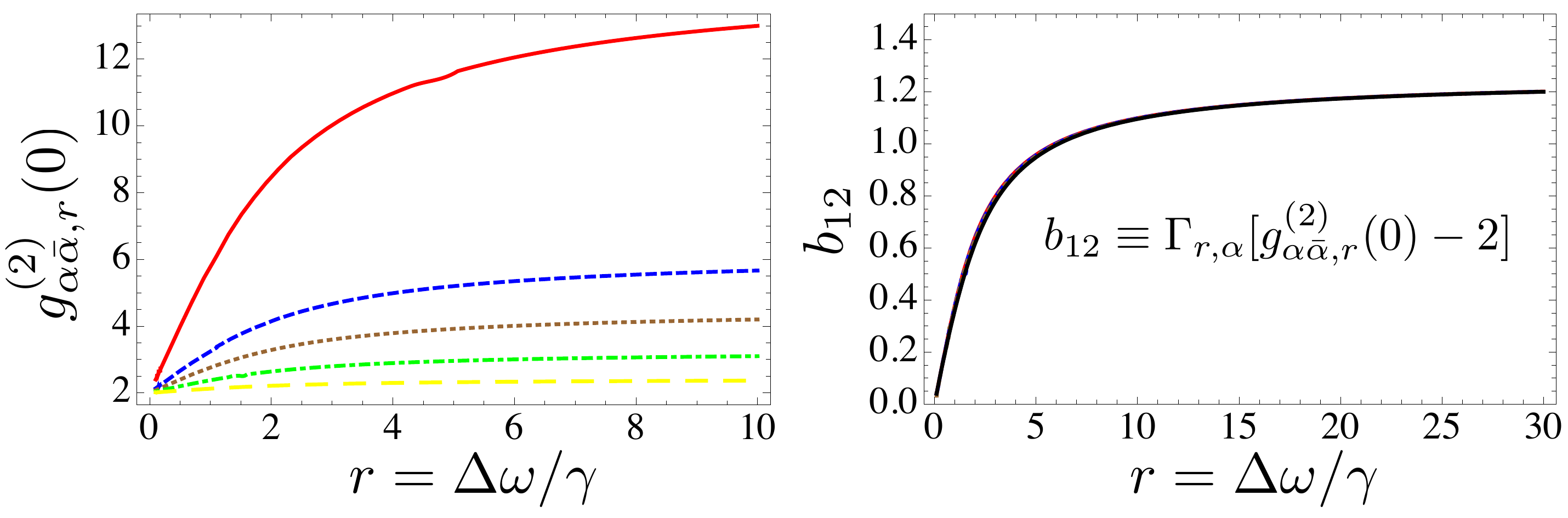}
\caption{Left: The cross-correlation factor $g^{(2)}_{\alpha\bar{\alpha},r}(0)$ as a function of the bandwidth $r=\Delta\omega/\gamma$ for various values of the rate $\Gamma$. The red, blue, brown, green, yellow curves correspond to $\Gamma=0.1$, $0.3$, $0.5$, $1$, and $3$, respectively. Right: The $b_{12}$ function in Eq.~\eqref{b12}, together with the approximate expression in Eq.~\eqref{b12app}. Here all curves, for which we used the same parameters as in the left plot,  lie on the same line,  and are fitted well by the approximate result.}
\label{cross_correlation_finite}
\end{center}
\end{figure}

Next we calculate the second-order correlation functions. We obtain:
\begin{align}
G^{(2)}_{\alpha\alpha,r}(0)&=\frac{\gamma^2\beta^2}{\pi}\left[\frac{\arctan\left(\frac{r}{1-\beta}\right)}{1-\beta}-\frac{\arctan\left(\frac{r}{1+\beta}\right)}{1+\beta}\right]^2\,,\\
G^{(2)}_{\alpha\bar{\alpha},r}(0)&=\frac{\gamma^2\beta^2}{\pi}\left[\frac{\arctan{\left(\frac{r}{1+\beta}\right)}}{(1+\beta)^2}+\frac{\arctan{\left(\frac{r}{1-\beta}\right)}}{(1-\beta)^2}\right]\,,
\end{align}
which results again in $g_{\alpha\alpha}^{(2)}(0)=2$, independent on the bandwidth $\Delta\omega$, while for the cross-correlation coefficient we obtain:  
\begin{align}
g_{\alpha\bar{\alpha},r}^{(2)}(0)&=1+\frac{\left[(1+\beta)\arctan{\left(\frac{r}{1-\beta}\right)}+(1-\beta)\arctan{\left(\frac{r}{1+\beta}\right)}\right]^2}{\left[(1+\beta)\arctan{\left(\frac{r}{1-\beta}\right)}-(1-\beta)\arctan{\left(\frac{r}{1+\beta}\right)}\right]^2}\,,
\end{align}
which instead  dependents on the bandwidth $\Delta\omega$.  Let us analyze this expression in more detail.  In experiments, one measures the photonic rate exiting the device, instead of the  bare emission rate $\beta$. Thus, we should calculate $g_{\alpha\bar{\alpha}}^{(2)}(0)$ as a function of $\Gamma_{r,\alpha}$.   Assuming the emission rate is such that the system is far below the instability threshold,  $\beta\ll1$, we found the general form:
\begin{equation}
g^{(2)}_{\alpha\bar{\alpha},r}(0)=a_{12}+\frac{b_{12}}{\Gamma_{\alpha}}\,,
\label{b12}
\end{equation}
with $a_{12}\approx2$ for $\beta<1$ and all values of $r$, while $b_{12}$ is given by
\begin{equation}
b_{12}\approx\sqrt{\frac{2}{\pi}}\left[\arctan{r}-\frac{r}{2+r^2}\right]\,.
\label{b12app}
\end{equation}
In  Fig.~\ref{cross_correlation_finite} we show the dependence of $g^{(2)}_{\alpha\bar{\alpha},r}(0)$ (left) and $b_{12}$ (right) on $r$ for different values of the measured rate $\Gamma$. We see that $b_{12}$ increases monotonically with increasing $\Delta\omega$, and that it saturates to $b_{12}^{\rm max}=\sqrt{\pi/2}$.  To conclude the zero-time delay discussion, we see that the input-output results match the our findings from the density matrix approach, but that considering a finite bandwidth affects the correlation function which shows less bunching (i.e. less correlated emission) as the bandwidth is decreased to smaller values than the natural bandwidths of the oscillators $\Gamma$.  

\begin{figure}[t]
\begin{center}
\includegraphics[width=0.99\linewidth]{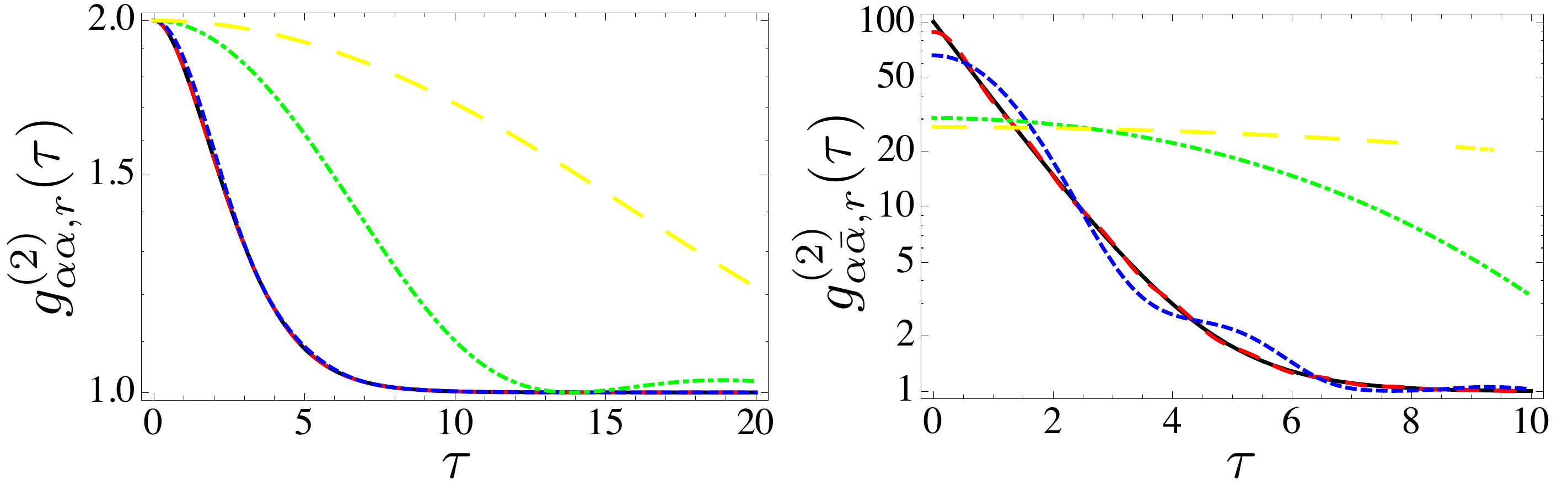}
\caption{Left (right): The auto-correlation (cross-correlation) factor $g^{(2)}_{\alpha\alpha,r}(\tau)$ [$g^{(2)}_{\alpha\bar{\alpha},r}(\tau)$] as a function of the time delay $\tau$ for different values of  $r=\Delta\omega/\gamma$.  The black, red, blue, brown, green, yellow curves correspond to $r=\infty$, $3$, $1$, $0.5$, and $0.2$, respectively, and we choose $\Gamma=0.1$. }
\label{correlation_finite_time}
\end{center}
\end{figure}

For completeness, let us also investigate the  time-dependence of the second-order correlation functions for a finite bandwidth detection. They are defined as in Eq.~\eqref{finite_time}, with the second-order correlation functions and rate being calculated over a finite bandwidth. While even in this case there are possible analytical expressions for the $g_{\alpha\alpha',r}^{(2)}(\tau)$ functions, they are too lengthy and uninspiring to be displayed. However,  in the left (right)  plot in Fig.~\ref{correlation_finite_time} we show $g_{\alpha\alpha,r}^{(2)}(\tau)$ [$g_{\alpha\bar{\alpha},r}^{(2)}(\tau)$] as a function of the delay time $\tau$ for various values of $r$. We see that the finite bandwidth strongly modifies the decay of the correlation functions for $r<1$, which now shows correlations over a longer time scale, of the order $1/\Delta\omega$. Note that the effective linewidth of the oscillators in the presence of the JJ is of the order of $\gamma(1-\beta^2)$, which needs to be compared with $1/\Delta\omega$. The cross-correlation Fano factor is also modified, and for $\beta\ll1$ it reads $F^r_{12}=2+5\Gamma f\left(r\right) $, with 
\begin{equation}
f\left(r\right)=\frac{\pi[33r+40r^3+15r^5+15(1+r^2)^3\arctan{r}]}{30(1+r^2)[r+(1+r^2)\arctan{r}]^2}\,.
\end{equation}  
This function behaves as $f\left(r\right)\approx2\pi/5r$ ($f\left(r\right)\approx1+8/3\pi r^3$) for $r\rightarrow0$ ($r\rightarrow\infty$).

\section{Discussions and Conclusions}
\label{sec5}

In conclusion, we have studied the radiation emitted by a voltage biased  JJ into two $LC$ oscillators when the the bias is set such that $2eV=\hbar\omega_1+\hbar\omega_2$. We have employed both the density matrix approach to study the cavity fields, and the input-output description to analyze the emitted photonic fluxes in the so called weak coupling regime characterized by an environmental impedance much smaller than $R_K$.   Specifically, we have calculated both the photon number and  the photonic correlations  (second-order coherence function) and showed, by proving that  a Cauchy-Schwarz inequality is violated, that the emitted radiation is non-classical. In order to analyze the effective dynamics of the oscillators,  we also calculated the time-dependence of the photonic correlations and showed that the their linewidth becomes narrower as the emission rate increases, signaling the approach to the  threshold instability.  We also briefly considered the effect of a finite bandwidth detection and showed that the qualitative features stay the same, but that the violation of the Cauchy-Schwarz inequality becomes less pronounced. In the future, it would be interesting to address also the strong coupling regime together with the region around the threshold instability, as well as  the full counting statistics of the emitted photons, as discussed  in Ref.~\onlinecite{Padurariu2012}.    

 \section{acknowledgments}
 We would like to gratefully thank Fabien Portier and Oliver Parlavecchio for  regular discussions motivating this work.
This work is supported by a public grant
from the ÒLaboratoire dÕExcellence Physics Atom Light MatterÓ
(LabEx PALM, reference: ANR-10-LABX-0039). 

{\it Note added}. During the final completion of this paper, we became aware of similar results obtained by A. Armour et al. arXiv:1503.00545 using the density matrix approach for the cavity fields.  %(Sec.~\ref{sec4} of this paper).

% for communicating their experimental results prior to publication and

%merlin.mbs 2010-03-15 4.21a (PWD, AO, DPC)

%\bibliography{Biblio}

\end{document}